\input{aipcheck}

\documentclass[,final] {aipproc}

\layoutstyle{6x9}

\newcommand{\be}{\begin{equation}}
\newcommand{\ee}{\end{equation}}
\newcommand{\bea}{\begin{eqnarray}}
\newcommand{\eea}{\end{eqnarray}}
\begin{document}

\title{ On the  determination of $\sigma$  from experimental  data on the  $\pi\pi$ isoscalar $S$-wave at low energies}


\author{Irinel  Caprini}{
  address={National Institute of Physics and Nuclear Engineering\\
POB MG 6, Bucharest, R-077125 Romania }}

\begin{abstract} We investigate the determination of the pole associated to $\sigma$ from $\pi\pi$ scattering data below the $K\bar{K}$ threshold, including the new  precise data from $K_{e4}$ decay reported recently by the NA48/2 Collaboration. Using a large class of analytic parametrizations based on expansions in powers of conformal variables, we obtain for the mass and width of $\sigma$ values  which  are consistent with those calculated recently using ChPT and Roy equations, but have larger theoretical uncertainties.

\end{abstract}

\maketitle


\section{Introduction}

 A precise determination of the mass and width of the $\sigma$ resonance  using Chiral Perturbation Theory (ChPT) and Roy equations was obtained recently in \cite{CCL}:
\be\label{CCL}
M_\sigma = 441^{+16}_{\,-8}\, {\rm MeV}, \quad  \Gamma_\sigma/2= 272^{\,\,+9}_{-12.5} \, {\rm MeV}.
\ee 
In the derivation of this result, the isoscalar $S$-wave is calculated at low energies, and also in the complex plane, from Roy equations, using  experimental data at high energies and theoretical results on $\pi\pi$ scattering \cite{ACGL,CGL}. So, unlike in the standard way of detecting resonances, no experimental data on the partial wave with the quantum numbers of the resonance at low energies were used as input.  The approach based on Roy equations is very suitable in this case, where  the  pole associated to the resonance is located far from the physical region and the experimental data at low energy are quite poor.

Recently \cite{NA48},  NA48/2 Collaboration measured the phase shift difference $\delta_0^0-\delta_1^1$ at low energies from  $K_{e4}$ 
 decay,  with a  precision much greater than that of the older experiments
 \cite{Rosselet, Pislak}. This revived the interest in 
the determination of the
 pole associated to $\sigma$  in the standard way, {\em i.e.} by the direct analytic extrapolation of the $\pi\pi$ scattering data.  In \cite{GMPY}
  the authors consider  a
  representation of the isoscalar $S$-wave $t_0^0(s)$ as  an expansion in powers of a conformal mapping variable, and predict the mass and width of $\sigma$ with an accuracy comparable to that quoted  in (\ref{CCL}).

In the present work we focus on the problem of systematic uncertainties within the approach proposed in \cite{GMPY}. By enlarging the class of admissible analytic functions considered in that work,  we determine the mass and width of $\sigma$ from experimental data  below the $K\bar K$ threshold, with a more realistic estimate of the uncertainties.

Below we give a summary of the results  obtained recently in this study. More details on the analysis and other results  will be presented in a forthcoming publication \cite{Caprini}.
\section{Analytic parametrizations of the amplitude}
1.  Writing the isoscalar  $S$-wave $t_0^0(s)$ as: 
\be\label{t00}
t_0^0(s)= \frac{1}{\psi(s)- i \rho(s)},\quad\quad \quad\rho(s)= \sqrt{1-4 M_\pi^2/s}\,,
\ee 
elastic unitarity implies that  $\psi(s)$ is analytic  in  the $s$-plane cut only for $s\le 0$ and  $s\ge 4 M_K^2$ (we neglect the inelasticity due to the 4$\pi$ channel below 1 GeV), except for a pole at the Adler zero $s_A$, where  $t_0^0(s_A)=0$. The  effective range approximation amounts to an expansion  of $\psi(s)$ in powers of $s$ near $s=4 M_\pi^2$. The domain of convergence can be enlarged by expanding in powers of a variable which  conformally maps  the holomorphy domain onto the interior of a  disk \cite{Confmap}. The function 
\be\label{wa}
 w(s,\alpha )=\frac{\sqrt{s}- \alpha \sqrt{4 M_K^2-s}}{\sqrt{s} + \alpha \sqrt{4 M_K^2-s}},
\ee
with $\alpha >0$  arbitrary, maps the $s$-plane cut along $s\le 0$ and $s\ge 4 M_K^2$ onto the unit disk  
$|w|<1$ in the complex plane $w= w(s,\alpha )$,  such that $w(4 M_K^2,\alpha )=1$ and $w(0, \alpha)=-1$.  In  \cite{GMPY} the authors adopt the expansion 
\be\label{psiYgen}
\psi(s)= \frac{M_\pi^2}{s-s_A} \left[ \frac{2 s_A} {M_\pi\sqrt{s}}+ B_0+B_1 w(s, \alpha) +B_2 w(s, \alpha)^2+\ldots\right],
\ee  
with the particular choice  $\alpha=1$. In Eq.(\ref{psiYgen}), the first term in parantheses  compensates the singularity of $\rho(s)$ at $s=0$ in the denominator of (\ref{t00}),  removing a ghost of $t_0^0(s)$ on the real axis which would appear otherwise. A slightly  different form was also used in  \cite{GMPY}, with an additional factor $(\mu_0^2-s)/\mu_0^2$ inserted in (\ref{psiYgen}). This parametrization, which displays the energy where the phase shift passes through $\pi/2$, is useful for fitting narrow resonances, but is not suitable for broad resonances like $\sigma$, and we shall not use it.
  
\begin{figure}\label{wgen}
\includegraphics[height=3.cm]{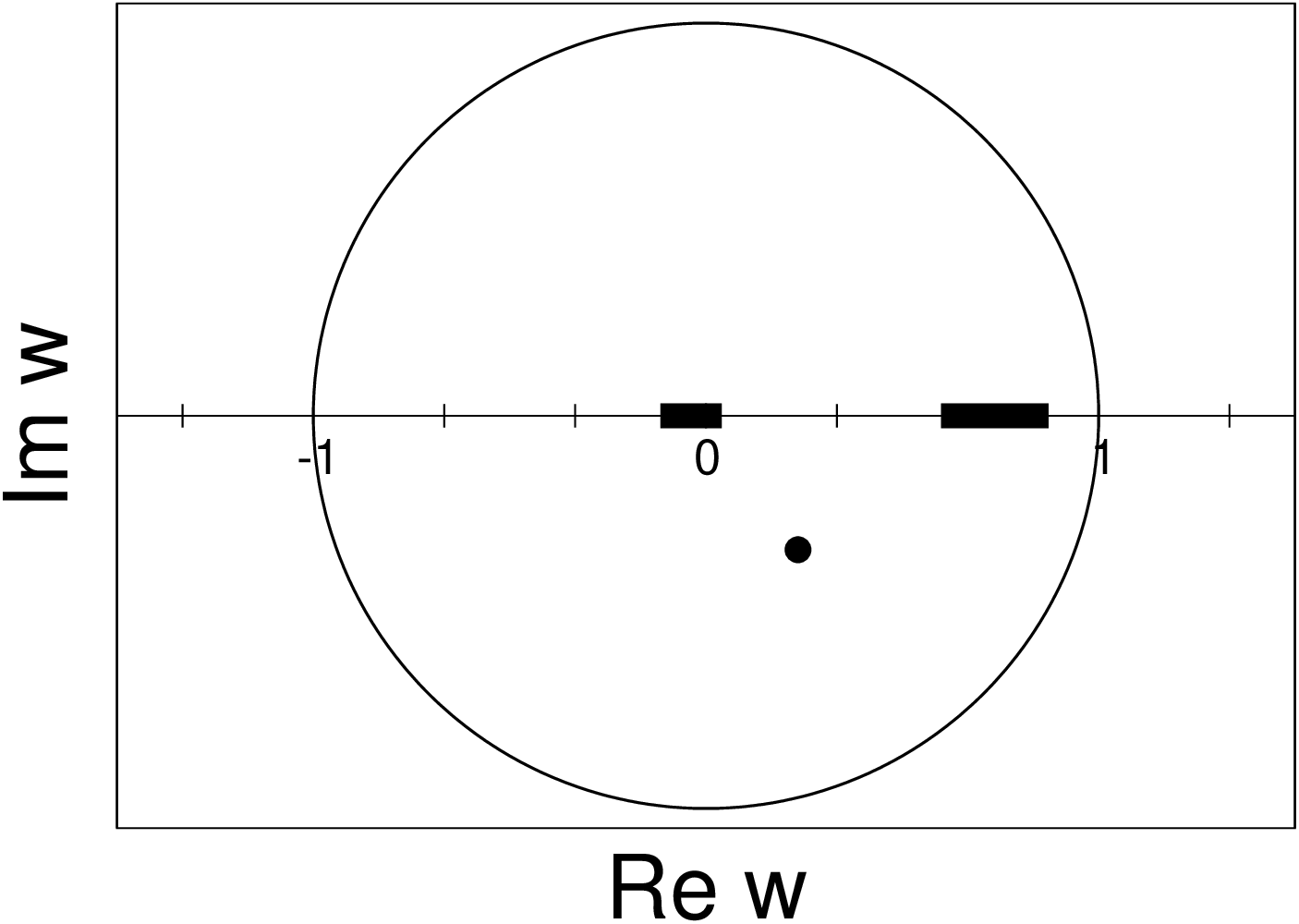}\hspace{0.5cm}\includegraphics[height=3.cm]{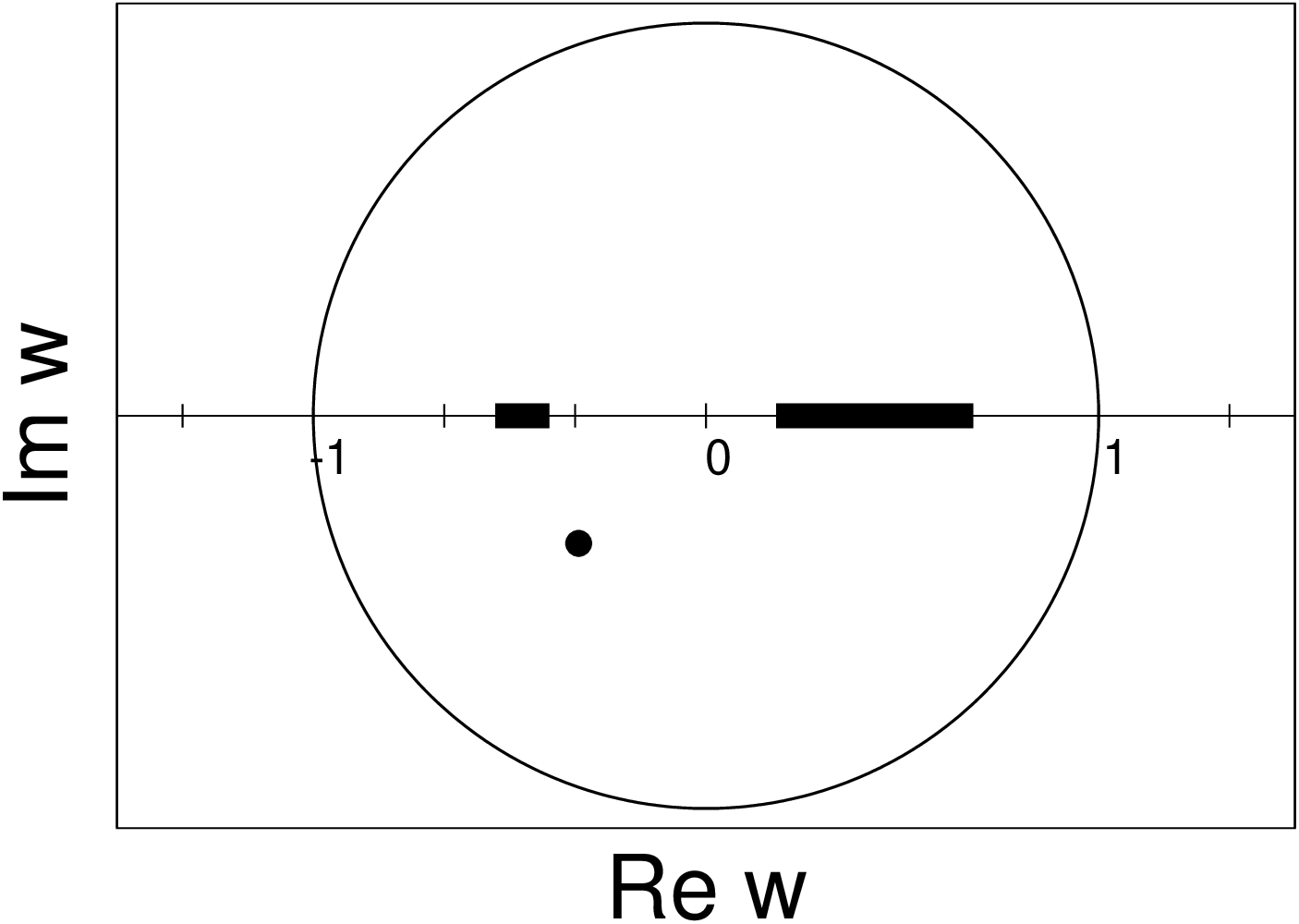}\hspace{0.5cm}\includegraphics[height=3.cm]{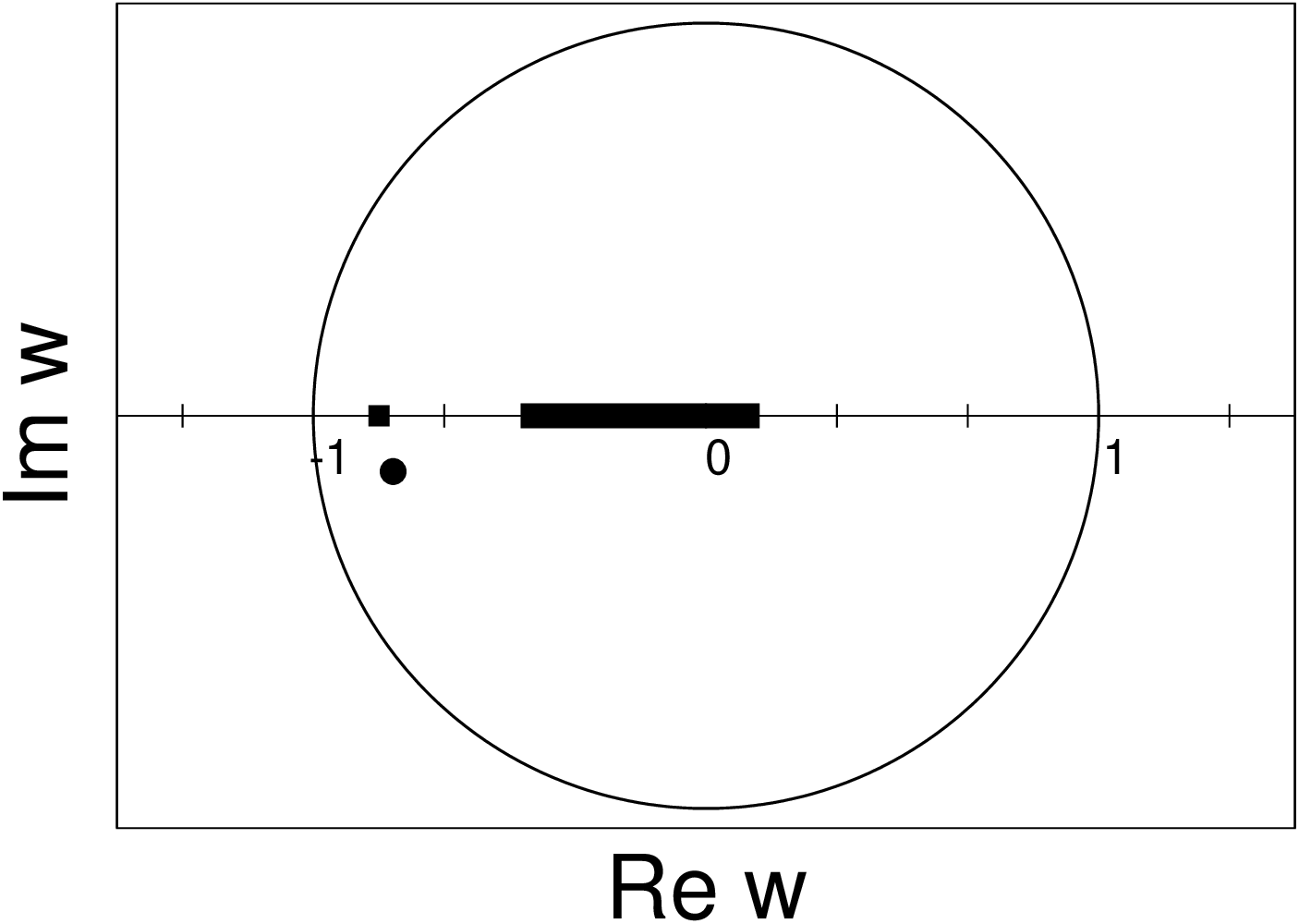}
\caption{The disk $|w|<1$ in the complex plane  $w=w(s,\alpha)$ defined in (\ref{wa}), for $\alpha=0.36$ (left), $\alpha=1$ (center)  and $\alpha=4$ (right). The thick segments indicate the regions where experimental data are available from $K_{e4}$ decay \cite{NA48}-\cite{Pislak} and the process $\pi N\to\pi\pi N$ (cf. the compilation given in \cite{PY 2005}), respectively; the circle shows  the $\sigma$ pole on the second Riemann sheet from \cite{CCL}.  }\end{figure}

A first generalization of  \cite{GMPY} is to expand $\psi(s)$ in powers of $w= w(s,\alpha )$ with an arbitrary $\alpha$, as in (\ref{psiYgen}).
 By varying $\alpha$, one changes the point mapped on the origin of the $w$-plane and the position of the intervals where experimental data are available. Some examples  are shown in Fig.~\ref{wgen}. 

\vskip0.3cm\noindent
2. An amplitude free of unphysical singularities can be alternatively obtained if
the term $i \rho(s)$ in (\ref{t00}) is replaced by a function which is  analytic in the $s$-plane cut along $s\ge 4 M_\pi^2$ and has  the imaginary part equal to $\rho(s)$ on the upper edge of the cut.  We consider for convenience the loop function  of ChPT:
 \be\label{J1}
J(s, M_\pi^2)) =\frac{1}{\pi}\left[2+\rho(s) \ln\left(\frac{\rho(s)-1}{1 + \rho(s)}\right)\right],
\ee
 which vanishes at the origin,  $J(0, M_\pi^2)=0$, and has  $\mbox{Im}\,J(s+i\epsilon, M_\pi^2)=\rho(s)$ for $s\ge 4 M_\pi^2$. Writing  the partial wave as
\be\label{t001}
t_0^0(s)= \frac{1}{\psi_1(s) - J(s, M_\pi^2)},
\ee 
elastic unitarity implies that the function  $\psi_1(s)$ can be expanded  most generally as
\be\label{psi1}
\psi_1(s)= \frac{M_\pi^2}{s-s_A}\left[B_0+B_1\, w(s,\alpha )+B_2 \,w(s,\alpha )^2+\ldots \right].
\ee  

\vskip0.3cm\noindent
3. Other parametrizations of $t_0^0(s)$ are obtained by expressing the $S$-matrix element as a product
\be\label{DalitzTuan}
S_0^0(s)= S_{\rm rest}(s) S_{f_0}(s),
\ee
where each factor satisfies elastic  unitarity ($|S_{\rm rest}(s)|=| S_{f_0}(s)|=1$) below the $K\bar K$ threshold.  It is convenient to parametrize the amplitude $t_{f_0}(s)$ associated to $S_{f_0}(s)$ as 
\be\label{tf0}
 t_{f_0}(s)= \frac{k_1 s } {\kappa -s- k_1 s \,J(s, M_\pi^2) -(k_2+k_3 s) J(s, M_K^2)},\ee
where $J(s, M_\pi^2)$ is defined in (\ref{J1}) and $J(s, M_K^2)$ is obtained by replacing $M_\pi$ with $M_K$. We note that with the choice
\be\label{kappaCohen}
 \kappa = 1.01, \quad  k_1 = 0.08, \quad k_2 = -1.09, \quad k_3 = 1.16,
\ee
the modulus of  $S_{f_0}(s)$ above the $K\bar K$ threshold is close to the elasticity $\eta_0^0(s)$ measured in \cite{Cohen}, while for
\be\label{kappaBES}
 \kappa =  1.15\,(1.41), \quad  k_1 = 0.11\,(0.24), \quad k_2 =  0.39\,(-0.73), \quad k_3 = 0.03\, (1.72),
\ee
it   follows the upper (lower) edge of the experimental band of $\eta_0^0(s)$  measured in \cite{BESf0}. 

 In our fits, both the phase shift and the elasticity  above the $K\bar K$ threshold are left free. The fact that we take a specific form for  $t_{f_0}(s)$ is not a limitation, since $t_0^0(s)$ contains an additional term,  $t_{\rm rest}(s)$, related to the factor $S_{\rm rest}(s)$ in (\ref{DalitzTuan}).  Elastic unitarity implies that  $t_{\rm rest}(s)$  can be parametrized, as in (\ref{t00}), in terms of a function $\psi_{\rm rest}(s)$ expanded as:
\be\label{psirest}
\psi_{\rm rest}(s)= \frac{M_\pi^2}{s-s_1} \left[ \frac{2 s_1} {M_\pi\sqrt{s}}+ B_0+B_1 w(s, \alpha) +B_2 w(s, \alpha)^2+\ldots\right],
\ee  
where $s_1$ is close to the Adler zero $s_A$ \cite{Caprini}. Admissible parametrizations are obtained also by replacing in this expansion  $w(s,\alpha )$ by  the variable $w_1(s,\alpha )= (\sqrt{s} - \alpha)/(\sqrt{s}+ \alpha)$,
which maps the $s$-plane cut only for $s\le 0$ onto the unit disk $|w_1(s,\alpha)|<1$. 

\section{Results and Discussion}
We used  the parametrizations described above for fitting the low energy data on the phase shift $\delta_0^0(s)$. In our analysis,
 the positive number $\alpha$ defining the conformal mappings  and the parameters $\kappa$ and $k_i$ appearing in (\ref{tf0})-(\ref{kappaBES}) represent the input which defines an admissible class.
In each class, the coefficients  $B_i$ of the expansion in powers of the conformal variable are  free. They are determined by fitting the low energy data.

We consider first the data on the difference  $\delta_0^0-\delta_1^1$ measured below 0.4 GeV from $K_{e4}$ decay \cite{NA48}-\cite{Pislak}. The $P$-wave 
is known with precision in this energy range \cite{CGL, PY 2005}, allowing an accurate extraction of phase shift of the $S$-wave. As in \cite{CCL}-\cite{CGL}, we work in the limit of exact isospin symmetry  and take for convenience, for both $M_\pi$ and $M_K$, the masses of the charged mesons. In order to obtain the strong  phase shift $\delta_0^0$, an isospin  correction  calculated recently in  \cite{Gasser} was
 subtracted from  the measured phase shift.  Following \cite{GMPY}, we increased the experimental error on the last point in \cite{Pislak}   by 50\%.  For the 10 data from the NA48/2 experiment we used the covariance matrix  published recently in \cite{NA48}.

\begin{figure}\label{Ke4}
  \includegraphics[height=0.22\textheight]{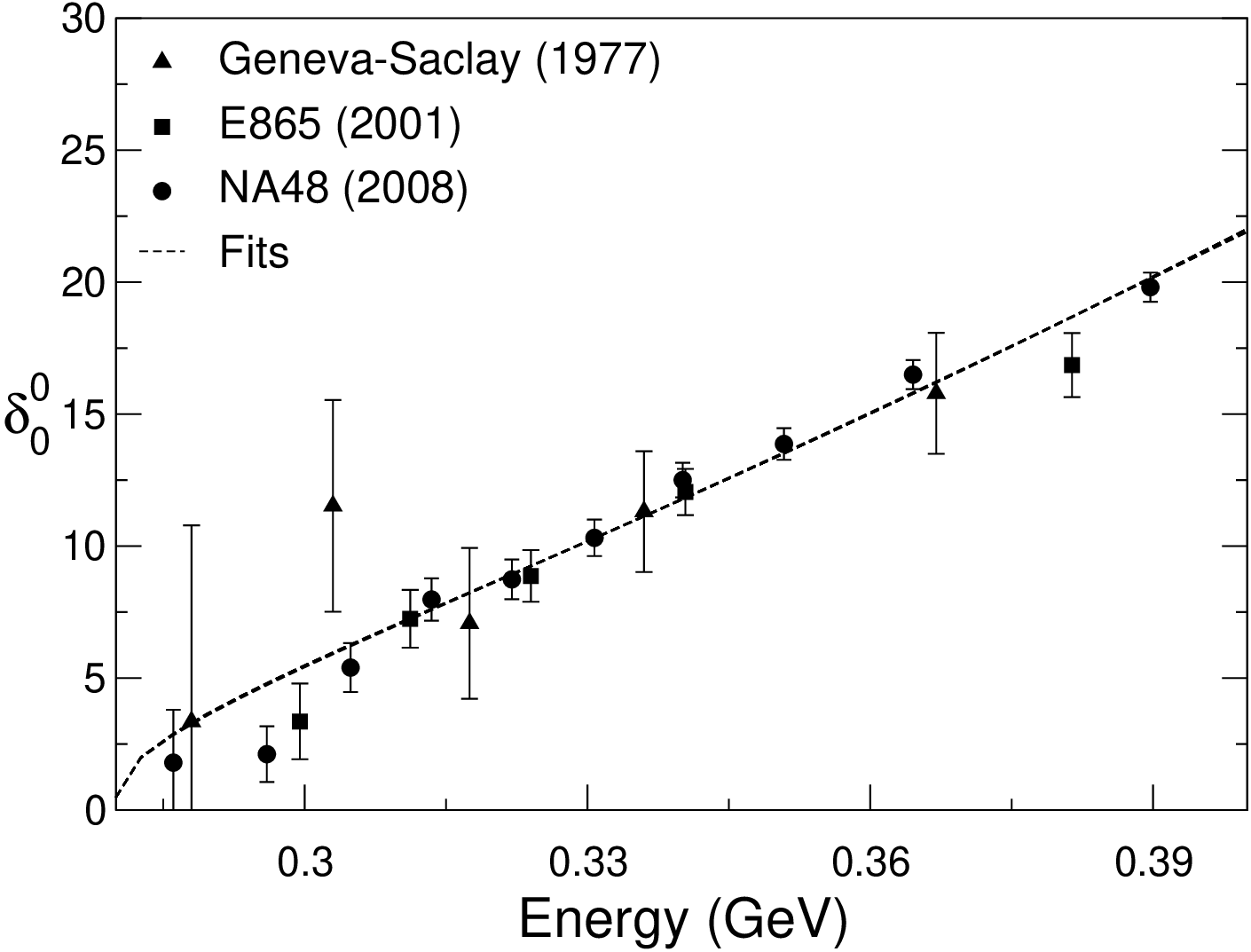}\hspace{0.4cm} \includegraphics[height=0.22\textheight]{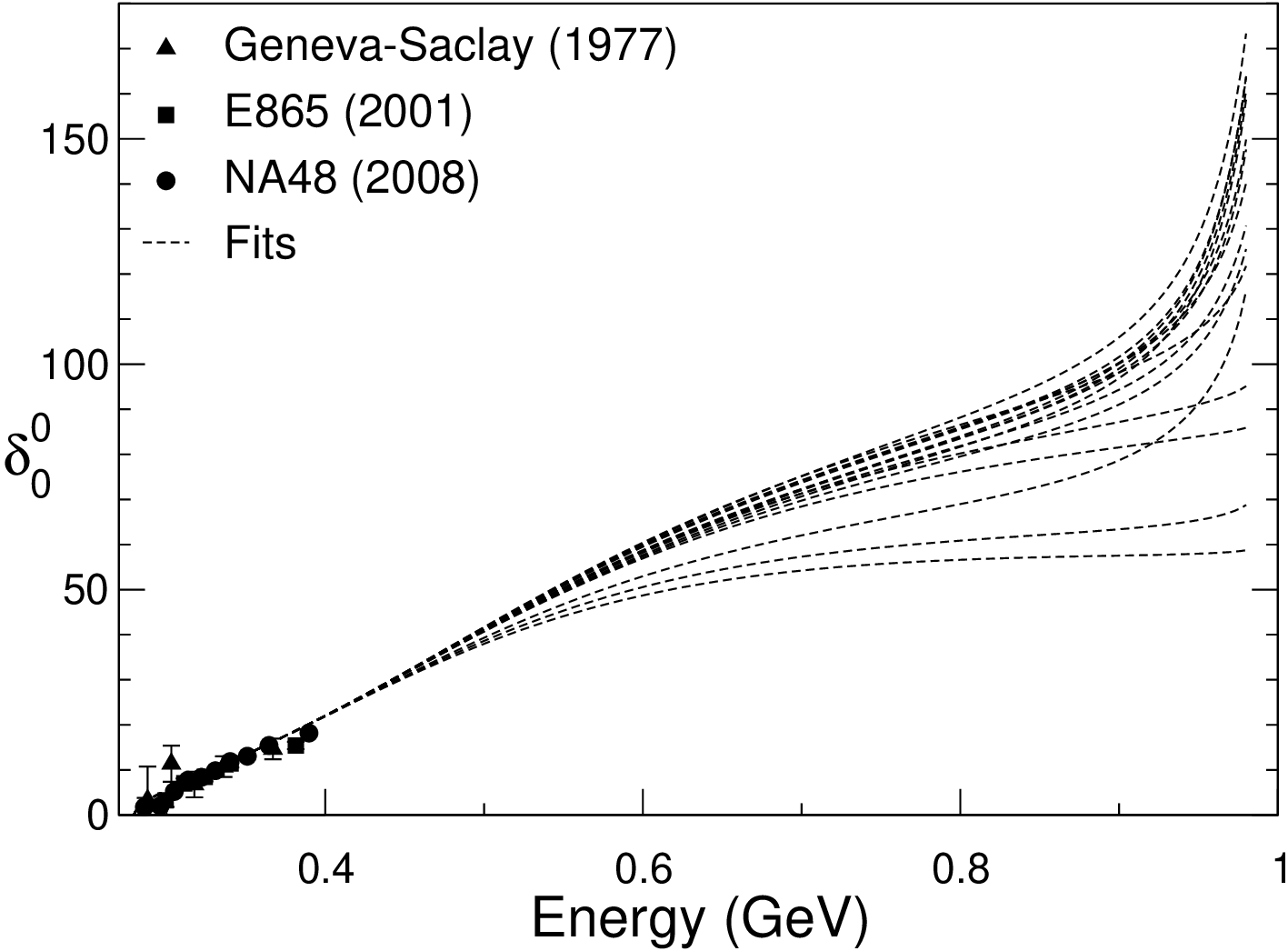}
  \caption{Left: phase shift  $\delta_0^0$ derived from $K_{e4}$ decay, fitted with the 16 parametrizations described in the text. Right: extrapolation of the parametrizations above the experimental range.}\vspace{0.4cm}
\end{figure}

We investigated 16 admissible parametrizations: the first 3  are based on Eqs. (\ref{t00}) and (\ref{psiYgen}) with $\alpha=1$, $\alpha=0.36$ and  $\alpha=4$, respectively, the next 3  are  based on Eqs. (\ref{t001}) and (\ref{psi1}) with the same choices of $\alpha$, and the last 10
 parametrizations are based on Eqs. (\ref{DalitzTuan})-(\ref{psirest})
 with various conformal mappings \cite{Caprini}. In each case the Adler zero $s_A$ was allowed to vary between 0.4 $M_\pi^2$ and 0.6 $M_\pi^2$. Using 2 free parameters, $B_0$ and $B_1$, for 21 points, the optimal values of $\chi^2$ for the 16 parametrizations are:  21.7, 21.5, 21.9, 20.9, 21.2, 20.6, 21.5, 21.7, 
21.8, 21.6, 21.8, 21.5, 21.7, 21.8, 21.6, 21.4 (they decrease by about 0.4 units if  the theoretical uncertainty of the isospin correction \cite{Gasser} is taken into account).

The quality of the fits is seen in Fig.~\ref{Ke4}, where the experimental points are obtained from  the data 
on $K_{e4}$ decay  \cite{NA48}-\cite{Pislak}  as discussed above.  Although the fits are almost indistinguishable in the experimental range, they exhibit large differences when extrapolated to higher energies, as seen in the right panel of Fig.~\ref{Ke4}. This illustrates the well-known  instability of analytic extrapolation \cite{Ciulli}. We note that the parametrization (\ref{t001}), especially with   the choices  $\alpha=0.36$ and $\alpha=1$ in (\ref{psi1}), leads to phase shifts which exhibit a plateau at low values. The  increase of $\delta_0^0$ required by the high energy data is  obtained, for instance, with $\alpha=4$ in Eq. (\ref{psiYgen}), or by using the $S$-matrix factorization (\ref{DalitzTuan}).

\begin{figure}\label{sigmamass}
  \includegraphics[height=0.25\textheight]{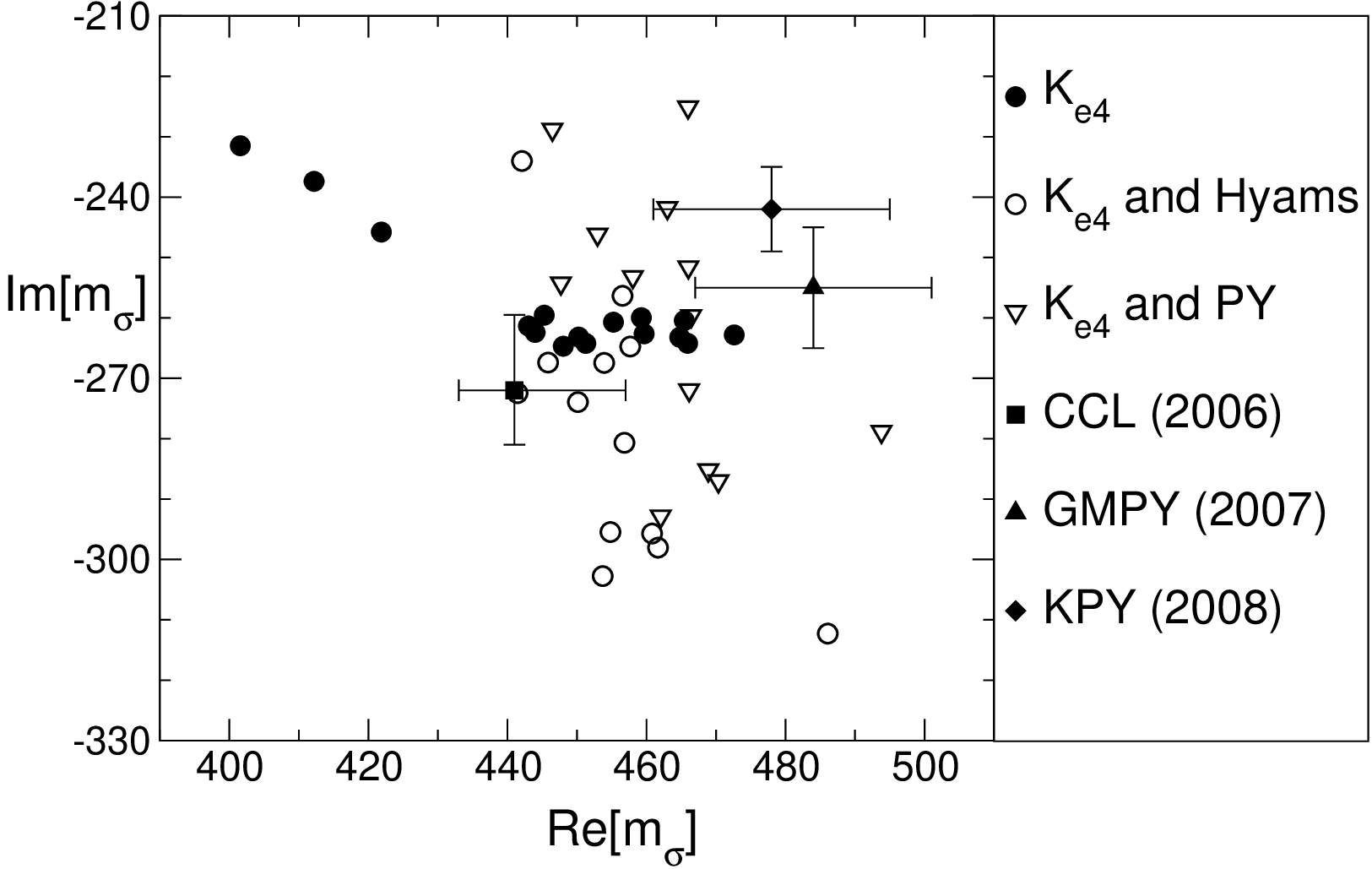}
   \caption{Positions of the $\sigma$-pole obtained by the analytic extrapolation of the parametrizations fitting the various sets of data, compared with  Refs. \cite{CCL}, \cite{GMPY} and \cite{KPY 2007} (from the last reference we show the value obtained with the isospin correction \cite{Gasser} included in the $K_{e4}$ data).}
\end{figure}

By extrapolating to the second Riemann sheet of the $s$-plane, we obtain the positions of the $\sigma$ pole  shown in Fig.~\ref{sigmamass}, where $m_\sigma=\sqrt{s_\sigma}= M_\sigma -i\Gamma_\sigma/2$. The three isolated points, with small values for both the mass and width, correspond to the fits based on the parametrization (\ref{t001}) mentioned above. Taking the average of the 16 admissible values of $M_\sigma$ and  $\Gamma_\sigma$  we obtain
\be\label{sigmaKe4}
M_\sigma=447 \pm 6\, \mbox{(stat)} \,\,^{+25}_{-46} \,\mbox{(syst)} \,\, \mbox{MeV},\quad \Gamma_\sigma/2= 258 \pm 6\, \mbox{(stat)}\, \,^{+10}_{-26}\, \mbox{(syst)}\,\, \mbox{MeV},
\ee
where the systematic error covers the spread of the admissible fits, including also an error of about 4 MeV in  $M_\sigma$ and 3 MeV in $\Gamma_\sigma$,
produced by the uncertainty in $s_A$. 

\begin{figure}\label{delta00HyamsPY}
  \includegraphics[height=0.22\textheight]{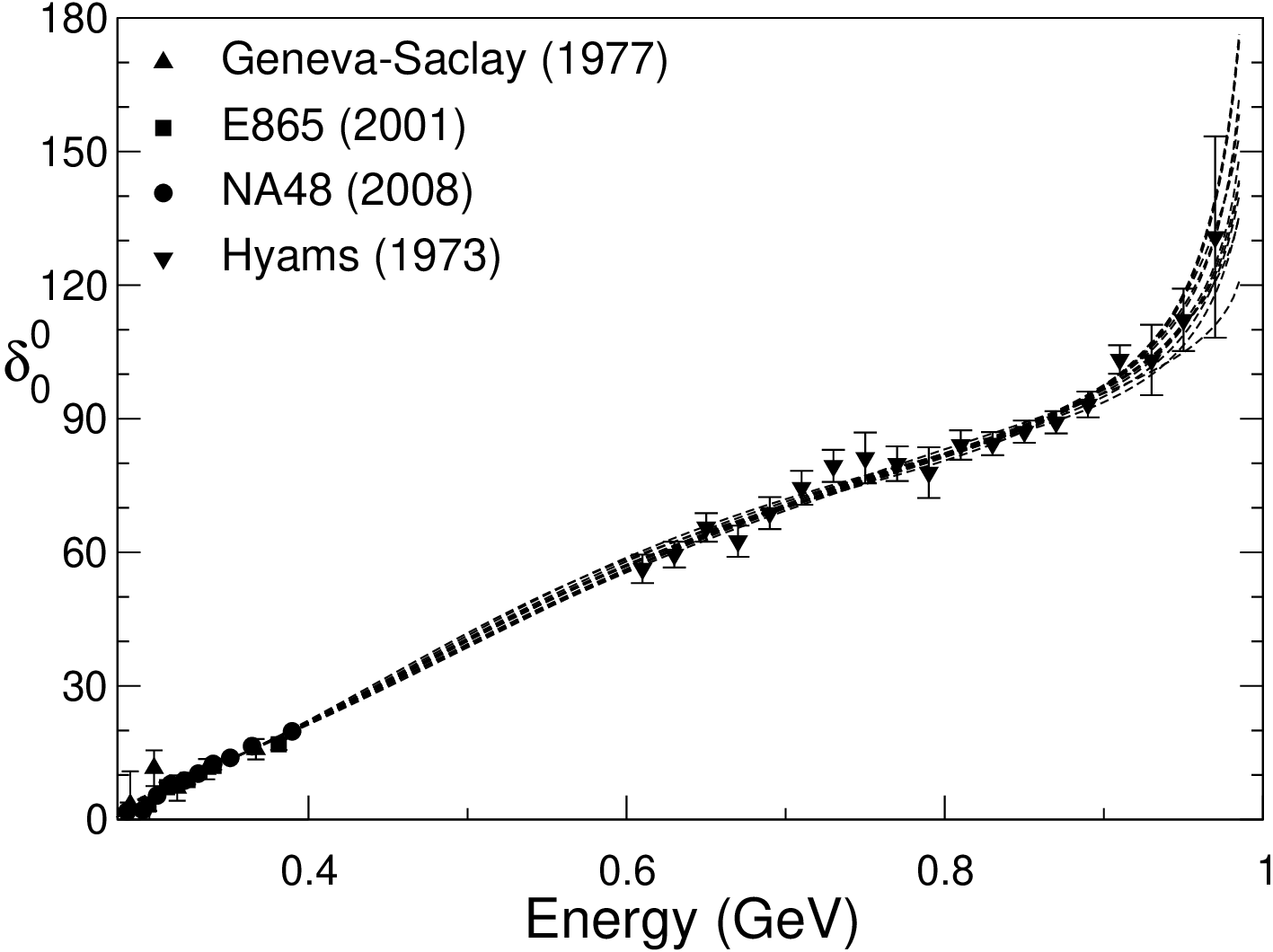}\hspace{0.5cm}
\includegraphics[height=0.22\textheight]{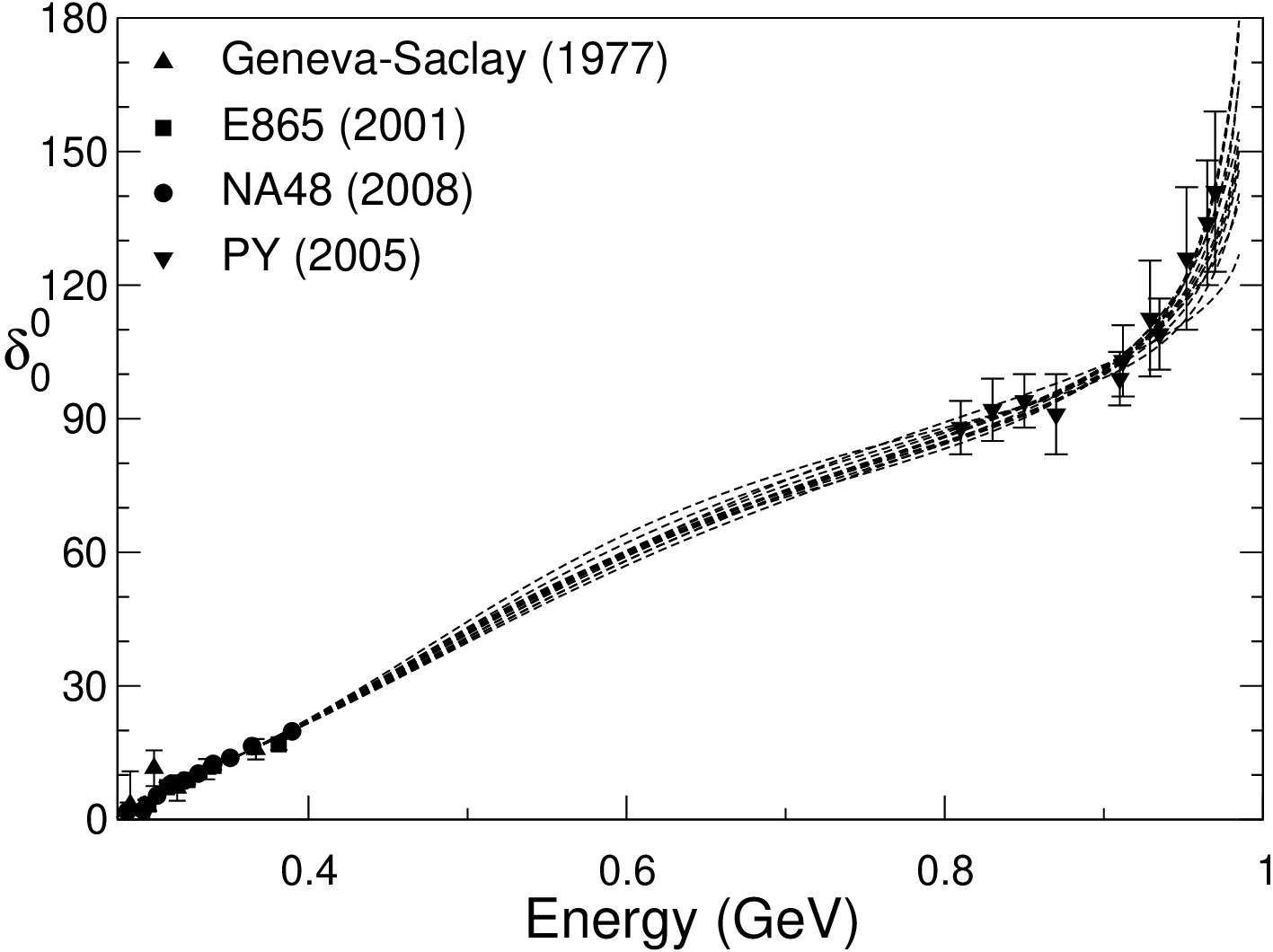} \caption{Left: fits of data in set I ($K_{e4}$ \cite{NA48}-\cite{Pislak}  plus CERN-Munich data \cite{Hyams} below the $K\bar K$ threshold). Right:  fits of data in set II ($K_{e4}$  plus a selection of data from $\pi N\to\pi\pi N$, given in  \cite{PY 2005}).}
\end{figure}

We can improve the description of $t_0^0(s)$  by including data on the phase shift at higher energies.  Two sets of data from $\pi N\to \pi\pi N$ were considered:  CERN-Munich data \cite{Hyams} consisting of 19 points below $4 M_K^2$,  and a collection of 11 data points given in Eq. (2.13) of \cite{PY 2005}.  Using 13  parametrizations similar to those described above \cite{Caprini}, we obtained,  with 2 or 3 parameters $B_i$, values for $\chi^2$ in the range (33, 38) for the 40 points of the set I, and  in the range (23, 29) for the 32 points of the set II. The quality of the fits is shown in Fig.~ \ref{delta00HyamsPY},  and the pole positions are given in Fig.~\ref{sigmamass}. The three isolated points present in the fits of  $K_{e4}$ data are no longer allowed. On the other hand, the  narrow range 
 of the widths $\Gamma_\sigma$, exhibited by the other fits of  $K_{e4}$ data, is now enlarged. This is due to the fact that the description of the $K_{e4}$ data,  measured by their contribution to the total $\chi^2$, is slightly worse than in the previous fits, and the various parametrizations are not as indistinguishable at low energies as in Fig. \ref{Ke4}.

Taking the average over the admissible parametrizations  we obtain, for the two sets:
\bea\label{sigma}
M_\sigma=455 \pm 6 \mbox{(stat)}^{+31}_{-13}\mbox{(syst)}\,\mbox{MeV},&\Gamma_\sigma/2= 277 \pm 6 \mbox{(stat)}^{+34}_{-43}\mbox{(syst)}\,\mbox{MeV}&{\rm (I)}\nonumber\\
M_\sigma=463 \pm 6\mbox{(stat)}^{+31}_{-17}\mbox{(syst)}\,\mbox{MeV},&\Gamma_\sigma/2= 259 \pm 6 \mbox{(stat)}^{+33}_{-34}\mbox{(syst)}\,\mbox{MeV}&{\rm (II}).
\eea
Alternatively, we can define the central values by selecting the fits with the lowest values of $\chi^2$ in each set. This procedure gives \cite{Caprini}:
\bea\label{sigmaopt}
M_\sigma=446 \pm 6 \mbox{(stat)}^{+40}_{-4}\mbox{(syst)}\,\mbox{MeV},&\Gamma_\sigma/2= 267 \pm 6 \mbox{(stat)}^{+44}_{-33}\mbox{(syst)}\,\mbox{MeV}&{\rm (I)}\nonumber\\
M_\sigma=458 \pm 6\mbox{(stat)}^{+36}_{-11}\mbox{(syst)}\,\mbox{MeV},&\Gamma_\sigma/2= 252 \pm 6 \mbox{(stat)}^{+39}_{-28}\mbox{(syst)}\,\mbox{MeV}&{\rm (II}).
\eea
 The comparison of (\ref{sigma}) and  (\ref{sigmaopt}) with (\ref{CCL}) shows that the $\sigma$-pole found by the analytic extrapolation of the low energy   data on the  $\pi\pi$  isoscalar $S$-wave  is consistent with the predictions of ChPT and  Roy equations. However, the theoretical uncertainties are now larger, since the differences between the various parametrizations of the partial wave  are amplified by the  extrapolation from the physical region to a distant point in the complex plane. In the method based on Roy equations \cite{CCL}, the instability of the extrapolation is tamed by using additional information on  the $\pi\pi$ amplitude.
  
\vspace{-0.2cm}
\begin{theacknowledgments}\vspace{-0.2cm}
I thank the organizers for this very nice and interesting meeting,
 and G. Colangelo, J. Gasser and H. Leutwyler for useful discussions and 
 comments.   This work was supported  
by the  Program CEEX of Romanian ANCS under Contract Nr.2-CEx06-11-92.
\end{theacknowledgments}

\end{document}